# A morphological indicator for comparing simulated cosmological scenarios with observations

E. Lega[1], A. Bijaoui[1], J.M. Alimi[2], and H.Scholl[3]

[1] Département C.E.R.G.A., Observatoire de la Côte d'Azur, B.P.229, 06304 Nice Cedex 4, France

[2] Laboratoire d'Astrophysique Extragalactique et de Cosmologie, U.R.A. CNRS 173, Observatoire de Paris-Meudon, 92195 Meudon, France

[3] Département G.D.Cassini, Observatoire de la Côte d'Azur, B.P.229, 06304 Nice Cedex 4, France

**Abstract.** We propose a morphological multi-scale analysis of large scale structures obtained by computer simulations and by observations. Structures are obtained at different scales by applying a wavelet transform on the observed and simulated data. Application of a segmentation algorithm allows a quantitative morphological description of the structures at each scale. The morphological parameter which we propose represents the deviation of a structure from sphericity. The dependence on scale of this simple parameter is shown to characterize, in simulations, the underlying cosmological model. We compare the HDM, CDM and mixed models with the CfA catalogue. This comparison favours a mixed model containing 65% of CDM, 30% of HDM, and 5% baryonic matter.

## 1. Introduction

In recent years, the amount of observational data has increased rapidly and we have now a picture for the distribution of galaxies on a large range of scales. At very large scales, the distribution of galaxies is homogeneous and isotropic while it is very inhomogeneous at smaller scales revealing a hierarchy of structures like voids, isolated galaxies, galaxy groups, galaxy clusters. Moreover, these systems of galaxies appear to be distributed on a connected network leaving most of the Universe empty (Joeveer and Einasto 1978, Einasto, Joeveer and Saar 1980, De Lapparent et al. 1986).

If the observed large scale structures are not just the accidental juxtaposition of smaller structures, the observed structures may contain clues to their origin and might be used to discriminate between different theories for structure formation. For example, in the models where the large scale structure is assumed to be formed by cosmological explosions (Ostriker & Cowie 1982), galaxies are situated on shells surrounding voids. On scales smaller than the void size, galaxies appear to be distributed on sheets. In hierarchical models, such as CDM, structures evolve by aggregation of matter into larger and larger clusters. Alignments of such clusters connected by low density filaments, first reported by Melott et al. (1983), appear as a consequence of the nonlinear phase of the evolution. These large scale structures depend on cosmological parameters like $\Omega_o$, $\lambda_o$ or on the presence of hot dark matter (MDM scenarios). Therefore we expect that the morphological details of structures as a function of scale vary from one model to another.

The first step to perform a morphological analysis of structures is, of course, the computerized detection of structures. Many methods have been proposed for extracting structured components from a noisy signal. They can be divided into two main classes: "local" and "global" methods. For the analysis of the large scale structure, for instance, the standard global methods are Fourier analysis, correlations analysis (Peebles 1980), multi-fractal description (Jones et al. 1988, Peebles 1989) and topological description using a genus quantity (Gott et al. 1986) . They yield general mean properties of the distribution of structures. The "local" methods like percolation (Shandarin 1983, Zel'dovich et al. 1982, more recently Klypin and Shandarin 1993), cluster analysis (Materne 1978, Geller et Huchra 1982) and smoothing and thresholding of data (Turner 1976) give the precise location and properties of each structure.

The wavelet transform is a local analysis at different scales which has the advantage, contrary to other methods, of being background independent. The detection of structures at a given scale by wavelet analysis is not affected

astro-ph/9510156   31 Oct 1995

by the existence of larger or smaller scale structures, like in the case of simple smoothing. Moreover, a statistically significant detection of structures is performed without any choice of parameters like for instance the choice of a threshold in the case of a density field . Finally, this multi-scale analysis allows to reveal a possible hierarchy in the structures.

Therefore, we introduce this method for an objective local analysis and we use a segmentation analysis in order to determine quantitatively the morphology of individual structures. Moreover, the statistical tools like genus and percolation analysis measure the connectivity of the distribution of matter but not the shape of the structures. Our morphological analysis is therefore complementary to other methods. The detection of structures at different scales is based on a new method: the combination of wavelet transform and segmentation, which appears particularly well adapted for the analysis of cosmological structures .

In section 2 we outline the wavelet transform and segmentation algorithms, and we define the morphological parameter. In section 3, we apply our analysis on the CfA catalogue and on simulations for the CDM and HDM scenarios.

## 2. Structure detection and characterization

### 2.1. The wavelet transform

The wavelet transform, introduced by Morlet, consists in the decomposition of a function $f(x)$ on a basis obtained by translation and dilation of a particular function, the so-called mother wavelet, which is localized in both physical and frequency space. For simplicity, we give definitions and basic properties of the wavelet transform for a one-dimensional function $f(x)$. Our applications are for three-dimensional functions. The wavelet transform of a one-dimensional, real, square-integrable function $f(x)$ is defined by:

$$w(a,b) = \frac{1}{\sqrt{a}} \int_{-\infty}^{+\infty} f(x)\psi^*(\frac{x-b}{a})dx \qquad (1)$$

where $a$ is the scale of the analysis and $b$ is the translation parameter corresponding to the position of the wavelet $\psi(x)$ ($\psi^*(x)$ is the complex conjugate of $\psi(x)$). $w(a,b)$ is called wavelet coefficient. For $a = 1$ and $b = 0$, the wavelet $\psi$ is called mother wavelet. The difference between Fourier analysis and wavelet analysis is the property of the latter to be invariant under dilation: the function $\psi(x)$ dilated or contracted maintains the same shape. The wavelet analysis is like a mathematical microscope for which $\psi(x)$ is the optics and $a$ the resolution. In order to reconstruct $f(x)$ from its wavelet coefficients the analyzing wavelet function $\psi(x)$ must satisfy the admissibility condition:

$$C_\psi = \int_0^{+\infty} |\hat{\psi}(\nu)|^2 \frac{d\nu}{\nu} < \infty \qquad (2)$$

where $\hat{\psi}(\nu)$ denotes the Fourier transform of $\psi(x)$. Therefore, $\hat{\psi}(\nu)$ must be equal to zero at the origin. For differentiable functions $\psi(x)$, this implies that the integral of $\psi(x)$ must be zero and, therefore, that the wavelet analysis is not affected by the mean noise of the background. It remains now to choose the mother wavelet. This choice, obviously, depends on the purpose of the analysis and on the characteristics of the function $f(x)$. In the case of a set of discrete points with the purpose to identify structures on different scales, the so-called isotropical Mexican hat (Laplacian of a Gaussian) is known to be a suitable choice for the mother wavelet (Slezak et al. 1990). However, the computational costs with such a mother wavelet are very high, in particular for analyzing large data sets resulting from N-body simulations. The wavelet coefficients are obtained on a grid by evaluating equation (1) scale by scale at each data point. A computation at a given scale does not use the results of computations at previous scales. This problem can be overcome by using the so-called "à trous" algorithm (Holdschneider et al. 1989), which, moreover, allows to use a mother wavelet with shape and properties very similar to the Mexican Hat (Fig.1).

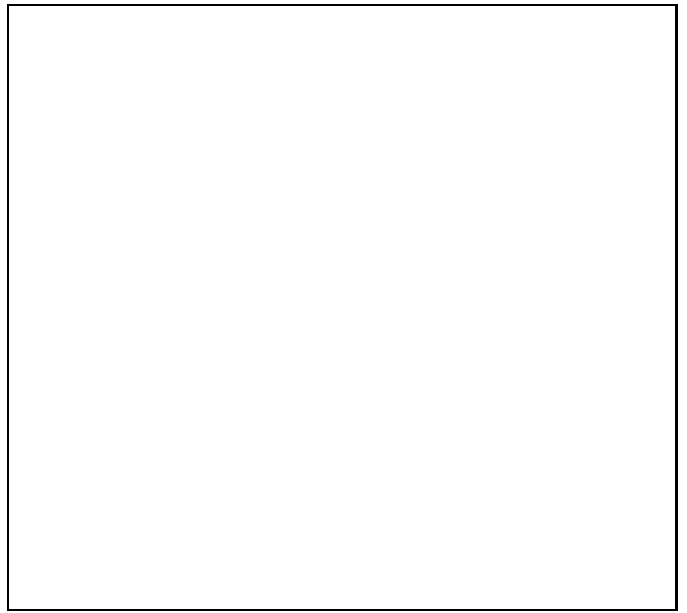

**Fig. 1.** a) The mexican hat, b) the mother wavelet for the "à trous" algorithm. The scaling function is the cubic B-spline.

The mother wavelet is then constructed from the cubic centered B-spline function $\phi(x)$ defined by

$$\phi(x) = \frac{|x-2|^3 - 4|x-1|^3 + 6|x|^3 - 4|x+1|^3 + |x+2|^3}{12} \quad (3)$$

and it has the following expression (Fig.1):

$$\psi(x) = \phi(x) - \frac{1}{2}\phi(\frac{x}{2}) \qquad (4)$$

The function $\phi$ is compact, regular up to the second order derivative and quasi-isotropic in the three-dimensional case ensuring the quasi-isotropic character of the wavelet analysis. A detailed description of the "a trous" algorithm is beyond the purpose of this paper. We like to refer, for instance, to E. Lega et al. (1995) for a detailed presentation of this algorithm and for its implementation on a data parallel computer (CM200).

## 2.2. The Segmentation

Segmentation (Rosenfeld 1969) is a widely used technique in the domain of image analysis. It corresponds to the determination of connected regions formed by pixels satisfying a given predicate $P$. Following Pavlidis (1977), we first define the predicate $P$. Given a grid $X$ with mesh points $j, l, m$ let $Y$ be a subset of $X$ containing one or more pixels, and let $M(j, l, m)$ be a matrix defined on the grid $X$. Then a predicate $P(Y)$ assigns the logical value true or false to Y depending only on the values of $M$ for the points of $Y$. The segmentation of a grid $X$ for a predicate $P(X)$ is a partition of $X$ into disjoint nonempty subsets $X_1$, $X_2$,...,$X_n$ such that:

1. each set $X_i$ is connected.
2. on each $X_i$ the predicate $P$ is true.

In our case the grid $X$ is the set of all mesh points and the $X_i$ are the structures to be identified. We call a set connected when any two points may be joined by a path along an axis of the grid. The matrix $M$ is the set of wavelet coefficients and the predicate $P$ is true for pixels on which the wavelet coefficients are larger than a given threshold (see Lega et al. for the implementation of this algorithm on a data parallel computer, CM-200).

## 2.3. The thresholding

We choose as predicate a thresholding for the wavelet coefficients. Since the mean value of the wavelet function is equal to zero, the wavelet transform yields coefficients equal to zero for a constant function. Consequently, the existence of structures at a given scale is connected to wavelet coefficients with a large absolute value at this scale. The statistical fluctuations in the spatial repartition lead to coefficients different from zero, even for a locally uniform distribution. We applied a classical decision rule to decide the significance of a structure by testing the probability $P(c)$ of a coefficient $c$ to be greater (for a positive value) than the observed value $C$. Let $\epsilon$ be a statistical level. If:

$$P(c > C) < \epsilon \tag{5}$$

we can say that this pixel in the wavelet space belongs to a structure at this level. The resulting segmentation depends on $\epsilon$. If its value is too large, artifacts are detected and the resulting morphological parameters are not correct. If its value is too small, many real structures are removed. We choose $\epsilon = 0.0001$, which corresponds to a good compromise between false alarms and misses.

The thresholds $C_p$ and $C_n$ are defined by:

$$P(c > C_p) = \epsilon \qquad \text{or} \qquad P(c < C_n) = \epsilon \tag{6}$$

Their values are determined by a procedure described by Slezak et al.(1993).

## 2.4. The Morphological Parameter

Structures are now individually identified. For the quantitative analysis of morphological properties of the system we introduce a shape parameter which describes the deviation from sphericity. This shape parameter $L$ is defined by:

$$L(a) = 36\pi \frac{V^2(a)}{S^3(a)} \tag{7}$$

where $V(a)$ and $S(a)$ are respectively the volume and the surface of a structure detected at the scale $a$. $L$ is nearly zero for very flattened or elongated structures (the limit $L = 0$ is for sheets or filamentary objects), while the maximum value (normalized to one) is obtained for spherical structures. We measure at each scale the mean deviation of shapes from sphericity by:

$$\langle L(a) \rangle = \sum_{i=1}^{N_{obj}} \frac{L_i(a)}{N_{obj}} \tag{8}$$

in order to study the variation of morphology with scale. $N_{obj}$ is the number of structures at scale $a$.

## 3. Application to observations and numerical simulations

### 3.1. Simple structures

We first apply our method on a simple example for the sake of illustration. A simple example consists of clusters of pixels which form on a larger scale spheres, sheets or filaments.

We consider in two large scale structures, a sheet and a filament, formed by groups of spheres (Fig.2).

The distance between the center of the spheres in each group is $d = 0.15$ while the distance between groups is $D = 0.3$. The only morphological difference between the filament and the sheet is at very large scale ($l \geq D$). All structures are embedded in a noisy background. Results are shown in Fig.3. The morphological parameter $\langle L(a) \rangle$ ranges from nearly 1. to 0.3 when the scale becomes comparable with the distance $d = 0.15$ between spheres. At this scale, we detect 3 elongated objects as indicated by the value of $\langle L(a) \rangle = 0.3$. At larger scales, $\langle L(a) \rangle$ grows due to the sphericity of the analyzing wavelet. When the scale is comparable to the characteristic group distance $D = 0.3$ we detect one object in both large scale structures. Their morphological difference is clearly detected (Fig.3) by our method.

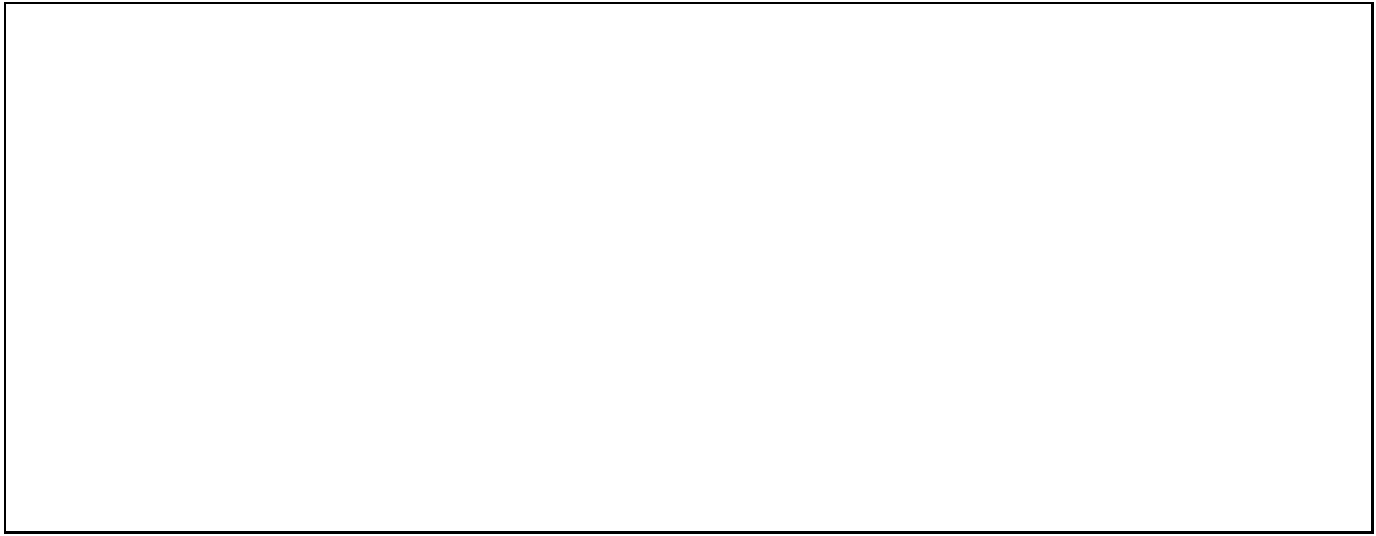

**Fig. 2.** Two dimensional projection of the distributions of objects described in section 3.1. a) Distribution of spheres which form a sheet at large scale. b) Distribution of spheres which form a filament at large scale.

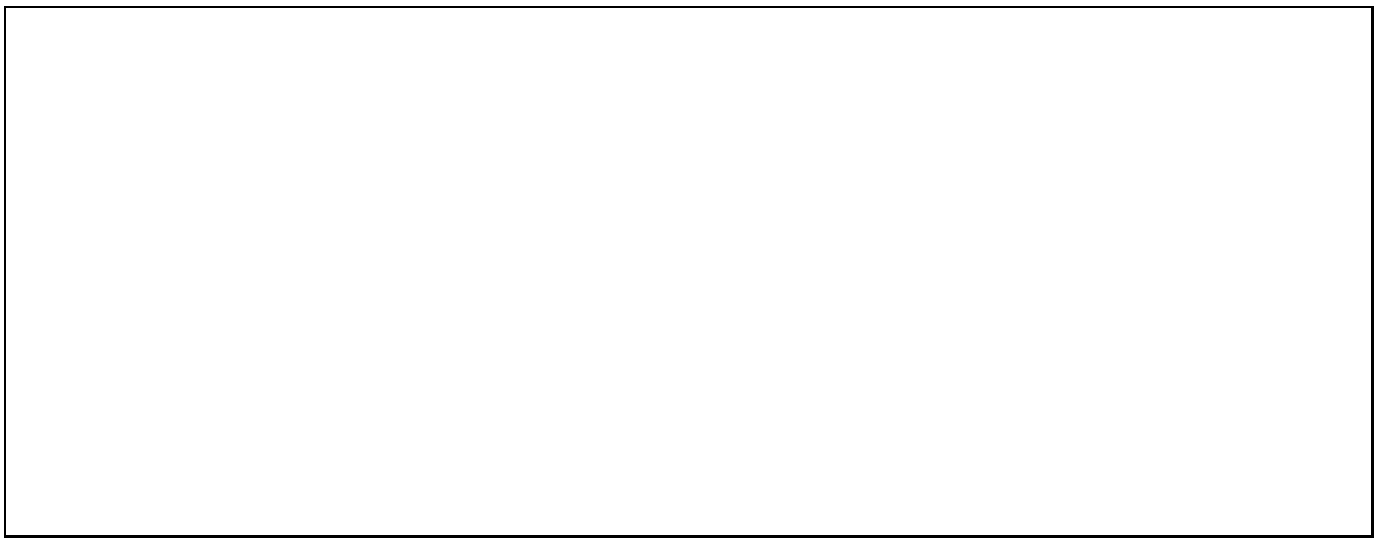

**Fig. 3.** a) Variation of $< L >$ with scale for the two distributions of objects, filament (solid line) and sheet (dashed line), described in section 3.1. b) Variation of the number of objects detected at different scales for the same two distributions.

### 3.2. The CfA catalogue

For the morphological analysis of the CfA catalogue, which represents the distribution of galaxies on a slice (de Lapparent Geller and Huchra 1986), we define the morphological parameter for the two dimensional case:

$$L(a) = 4\pi \frac{S(a)}{P^2(a)} \qquad (9)$$

where $S(a)$ and $P(a)$ are respectively the surface and the perimeter of an object detected at scale $a$. We applied the wavelet analysis to the CfA catalogue in the range $150Km/s \leq a \leq 850km/s$. The lower limit corresponds to the mean velocity of galaxies inside the groups. The upper value of $850km/s$ avoids artifacts in the measure of $L$ due to structures with edges extending over the borders of the catalogue. Moreover, the analysis is performed only on pixels at a distance of $d \geq 2^{a+1}$ from the border of the slice. This restriction avoids artificial boundary effects when applying the wavelet transform near the edges of the catalogue.

In order to avoid selection effects of the catalogue on the resulting morphological parameter, an analysis should be carried out preferably on a catalogue which is complete with respect to distance. Unfortunately, such a catalogue would contain too few galaxies. The CfA is complete up to a absolute magnitude of $M = -19$ but contains only 236 galaxies within this range. Such a small number would drastically reduce the scales for our analysis, and comparisons with simulations would not be meaningful. In order to enable comparisons, we introduced selection effects in the simulated data as well. Selection effects are thus in both samples which are compared. The results will be presented below.

Our results for the CfA catalogue are shown in Fig.4 and in Fig.5.

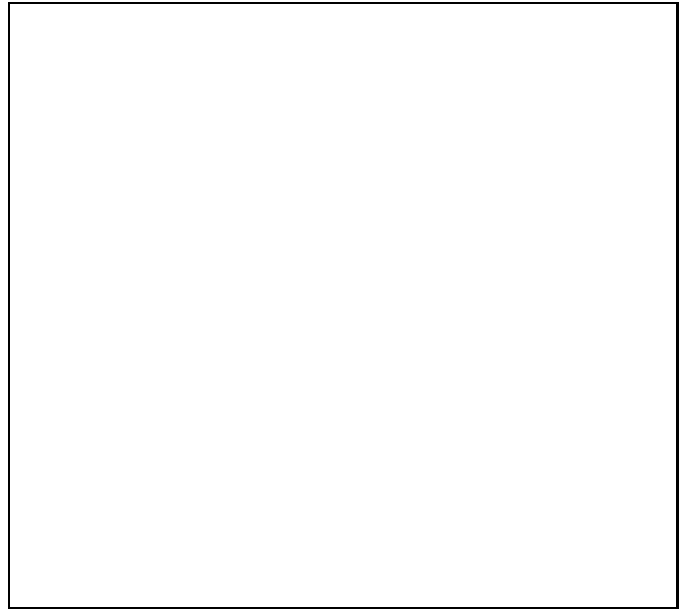

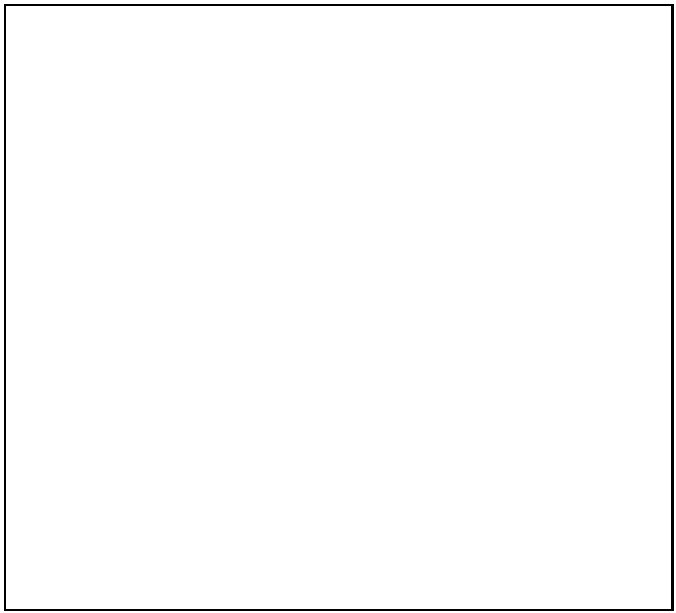

**Fig. 4.** Variation of $< L >$ with scale for the CfA catalogue.

The diminution of $\langle L \rangle$ and of the number of objects as a function of scale indicates a hierarchical distribution of structures. At small scales, circular shapes dominate. Shapes become more elongated with growing scale ($a \geq 500 km/s$).

### 3.3. Cosmological scenarios

We show in the following that our morphological parameter is very well suited to discriminate between the cosmological scenarios CDM (Bond & Slazay 1983, Davis et al.1985) and HDM (Centerella et al. 1988). In the HDM scenario, large scale primordial fluctuations dominate. The corresponding spectrum has a peak at the size of superclusters of galaxies. In the CDM scenario, on the other hand, primordial density fluctuations on subgalactic scales are dominant. We have analyzed the different resulting morphological large scale structures.

HDM and CDM scenarios were simulated by a Particle-Mesh code on a data parallel Connection machine with a resolution of $128^3$ (Alimi and Scholl 1993). Simulated "universes", of size $l = 192 Mpc\,h^{-1}$, with $h = 0.5$, containing $128^3$ particles and defined by the parameters $\Omega = 1$, $\Lambda = 0$, have been investigated for both scenarios.

We show in Fig.6 the results of the morphological analysis. Error bars are given by the variance on $\langle l \rangle$ obtained over 5 simulations for each scenario. It appears clearly that the cold scenario is characterized by structures of almost spherical shape slightly changing towards more elongated shapes when the scale exceeds $2 Mpc\,h^{-1}$. On the other hand, the hot scenario is characterized by elongated structures even at very small scales ($\simeq 1 Mpc\,h^{-1}$).

### 3.4. Two dimensional analysis of simulated scenarios

Since the CfA catalogue corresponds to a slice of the Universe, comparisons with simulated data should be made in 2-D rather than in 3-D. We extracted from the 3 dimensional simulations slices comparable to the CfA catalogue with respect to geometry and to size. The CfA catalogue has been compared with a CDM, HDM, and with a mixed scenario (MDM) (Holman et al. 1983) . The MDM model contains 65% of CDM, 30% of HDM, and 5% baryons as suggested by recent results about the compatibility of such a scenario with the anisotropy of the cosmic background radiation (Shaefer & Shafi 1992, Taylor & Robinson 1992). We see in Fig.7 the variation of the morphological parameter with scale for the three scenarios. Error bars are determined by the variance of $\langle L \rangle$ for 5 slices. We see a clear difference between the cold and hot scenario like in the 3-D case above. The error bars for the HDM case are larger than in the three dimensional case. This is due to the fact that even at small scales the hot scenario has elongated structures. Some elongated structures appear more circular in the 2-D sample due to the slice cutting. Due to this effect, we compared only the CDM and mixed scenarios with the CfA catalogue. These scenarios are quite well distinguished (Fig.7) up to a scale of $5 Mpc\, h^{-1}$.

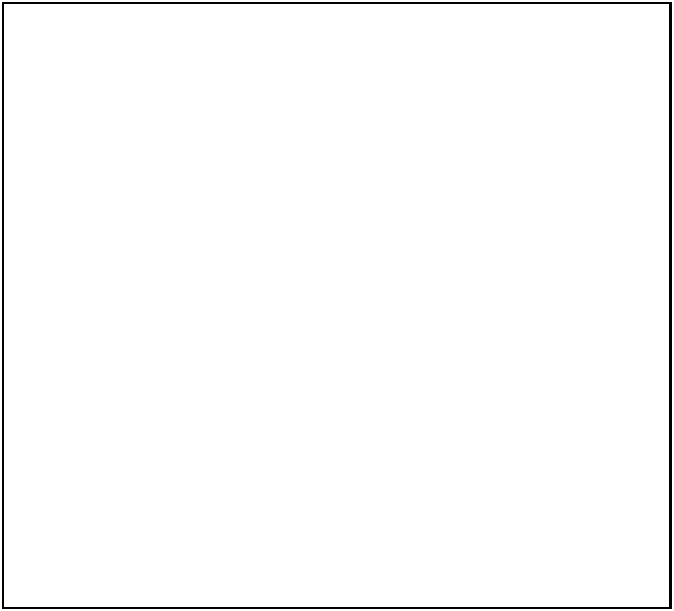

**Fig. 6.** a) Variation of $< L >$ with scale for the distribution of structures of the cosmological scenario CDM and HDM. Error bars are the standard deviation of $< L >$ obtained over 5 simulations for each scenario.

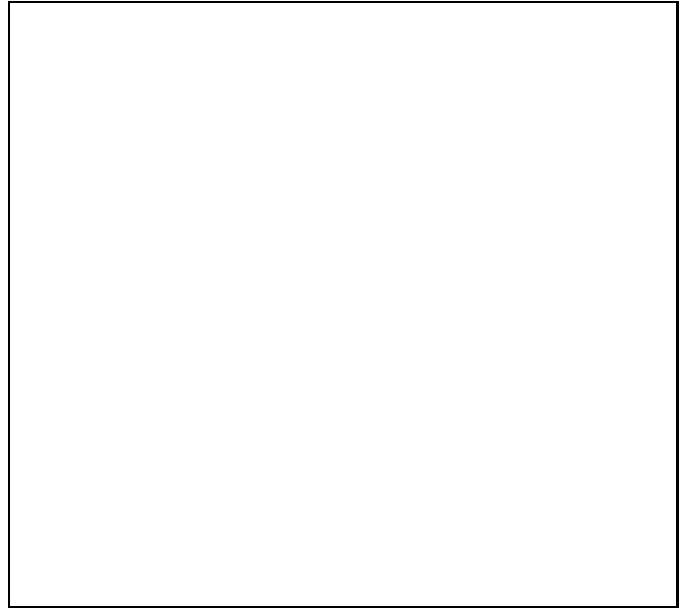

**Fig. 7.** Variation of $< L >$ with scale for catalogues comparables to the CfA one, extracted from the cosmological scenarii CDM, HDM and MDM.

### 3.5. Comparison between simulation and observations

For comparing the simulated scenarios with the CfA we have modeled in the simulated sample two major deficiencies of observations: observations are made in the redshift space and catalogues are effected by selection effects. The first deficiency, which causes the well known effect of the "Fingers of God", is particularly important in the case of a morphological analysis. Of course, this effect is more important in a scenario with a large velocity dispersion in groups of galaxies. According to Fig.8, the mean value and the variance of the velocity distribution of galaxies contained in a CDM slice are greater as compared to the MDM scenario. The effect of a "redshift space" elongation is therefore more important for the cold scenario than for the mixed one. This effect causes the strongest uncertainty for the morphological comparison between the two scenarios.

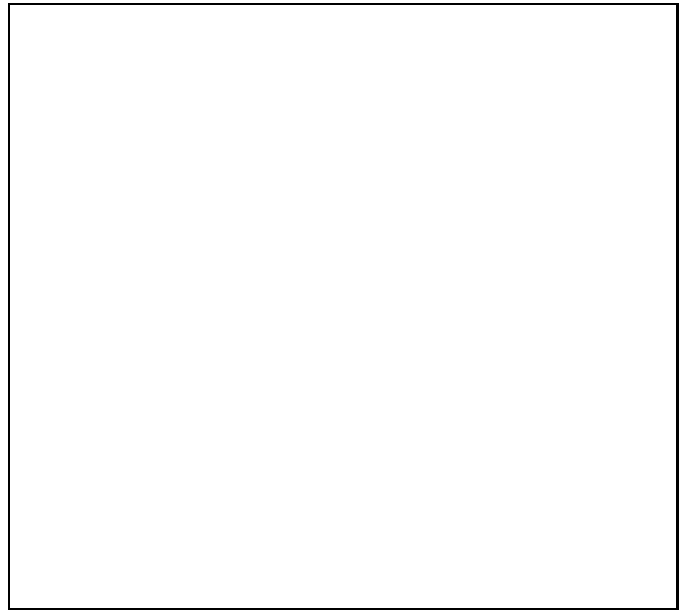

**Fig. 8.** Distribution of proper velocities of particles on a slice similar to the CfA one extracted from the cosmological scenarii CDM and MDM.

In order to take into account selection effects, we assigned magnitudes to each data point in the simulated sample in accordance with the luminosity function of the CfA (De Lapparent et al. 1989). Figures 9 and 10 give the respective morphological comparisons for the CfA with the cold and the mixed scenario. A chi square test yielded a value of $\chi^2 = 24.06$ for the cold and a value of $\chi^2 = 8.10$ for the mixed scenario. The mixed scenario is favoured by this analysis, but we stress the fact that the loss of information caused in particular by the restriction of the observational material to a 2-D slice makes comparisons between observations and simulations very difficult.

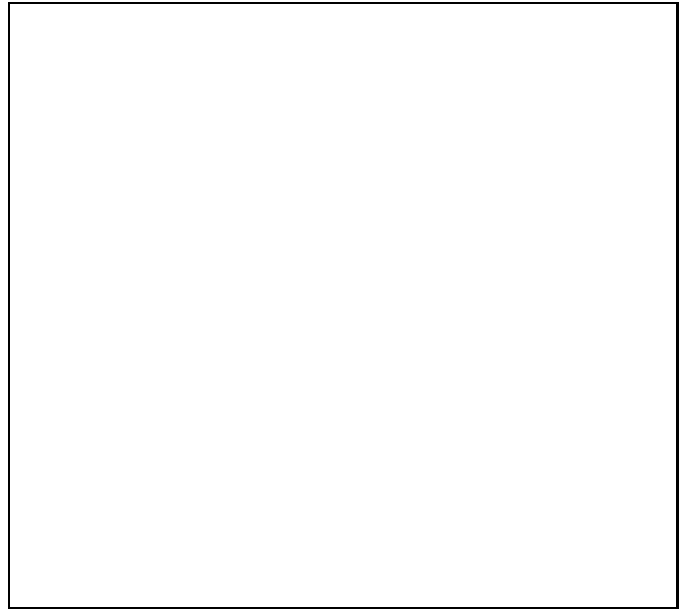

**Fig. 10.** Morphological comparaison between the CfA catalogue and the MDM scenario.

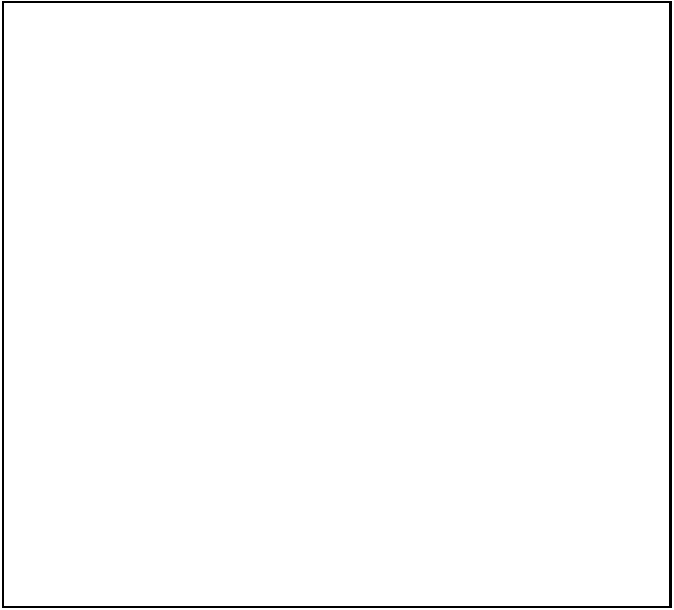

**Fig. 9.** Morphological comparison between the CfA catalogue and the CDM scenario.

## 4. Conclusions

We present a morphological method to compare simulated scenarios with observations. Our method consists of three steps:

- (i) Identification of structures at different scales by wavelet analysis
- (ii) Identification of pixels belonging to the same structure by segmentation
- (iii) Computation of a morphological parameter as a function of scale.

The identification of structures by wavelet analysis has the advantage to be an objective multi-scale method.

The basic idea is that the multi-scale comparison of the morphology of structures reveals the underlying scenario like, for instance, CDM or HDM. This idea has been tested successfully on simulations. We have chosen in these tests a very simple parameter to characterize the morphology of a structure, namely the deviation from spherical shape. This parameter appears to be sufficient to distinguish between a pure CDM and HDM scenario. We do not exclude that this parameter should be replaced for comparing other scenario like mixed ones by a parameter which gives more information about the morphology of a structure. We like to assess that the essential point of our method is the multi-scale comparison. It is therefore natural to use a multi-scale method like the wavelet analysis for the identification of structures at different scales. This method has also the advantage to be objective.

When comparing morphological structures obtained by simulations and by observations at different scales, we encountered difficulties which are not solely due to catalogue deficiencies but which are intrinsic and which set limits to such comparisons. The CfA catalogue has two major, catalogue dependent deficiencies for our purpose: observations are made in a 2-D slice and are affected by selection effects. The restriction to a 2-D slice may result in the detection of structures with an unrealistic morphology. For instance, an elongated structure may appear a 2-D slice as a circular structure. The second problem, selection effects may result in distortions of structures at larger distances. Both difficulties can be overcome by a corresponding increase of observational material. The size of the slice covered in the CfA catalogue should also be increased.

A more serious intrinsic problem arises from the fact that observations are made in the redshift space. Large velocity dispersions in groups of galaxies affect the reliability of resulting values for morphological parameters which characterize structures.

Our comparisons between simulated HDM, CDM and MDM scenario with the CfA catalogues is in agreement with a mixed model containing 65% of CDM, 30% of HDM, and 5% baryonic matter.

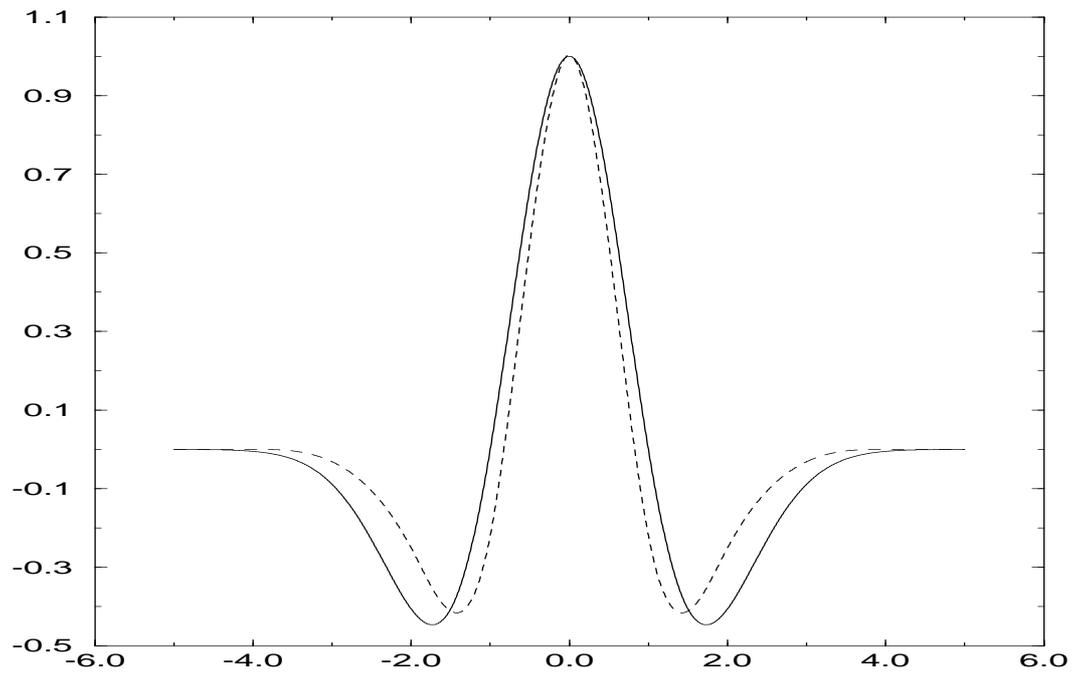

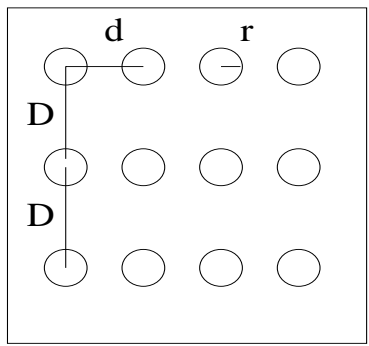

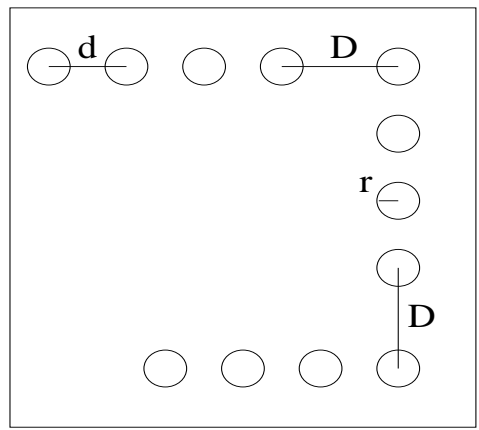

a                                    b

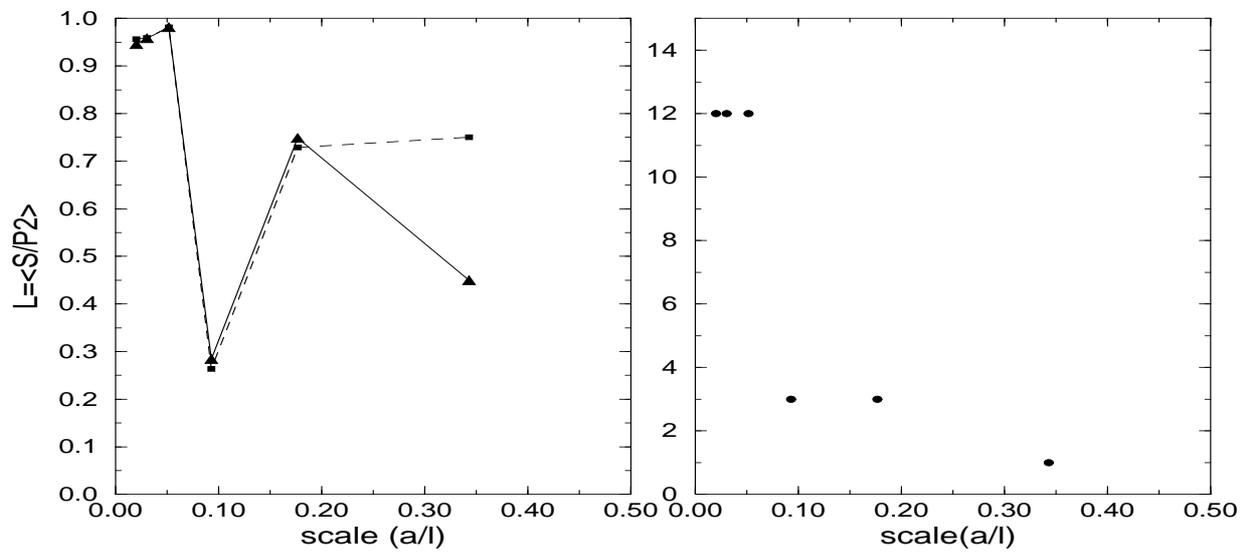

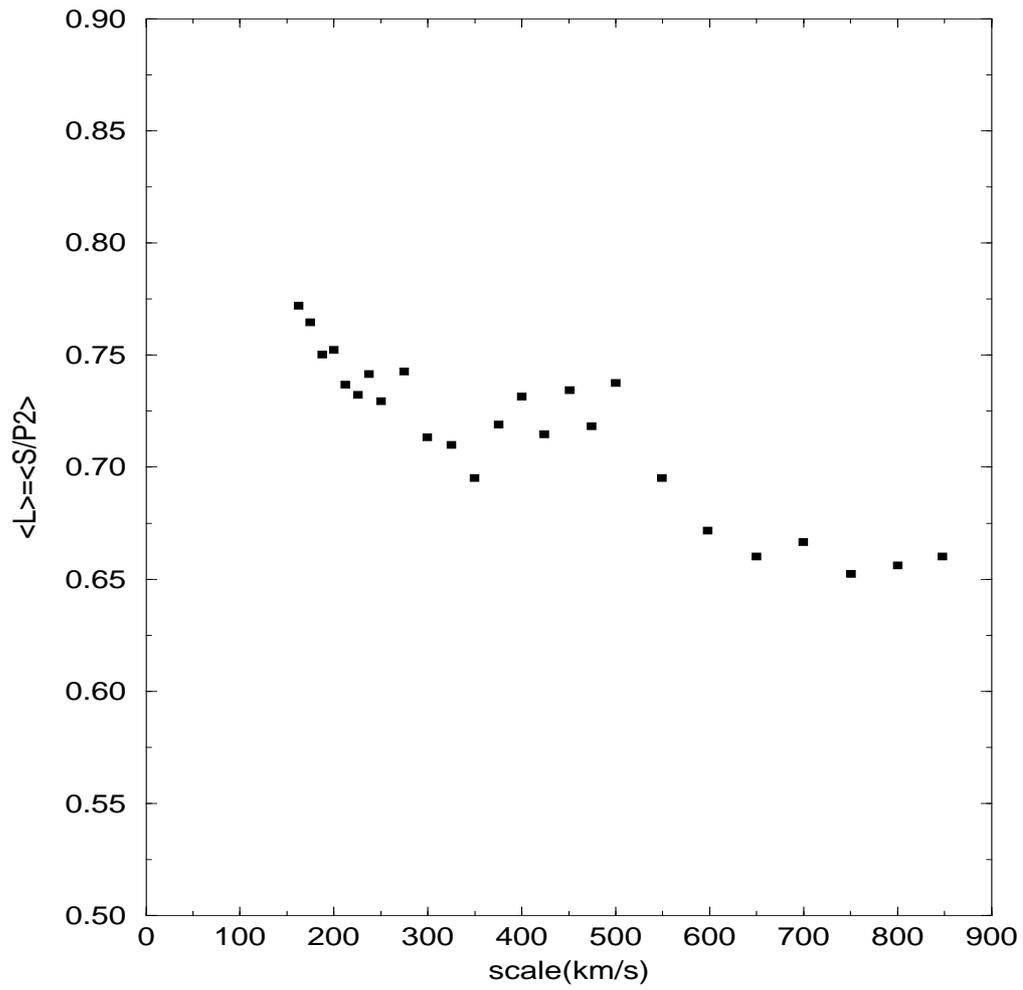

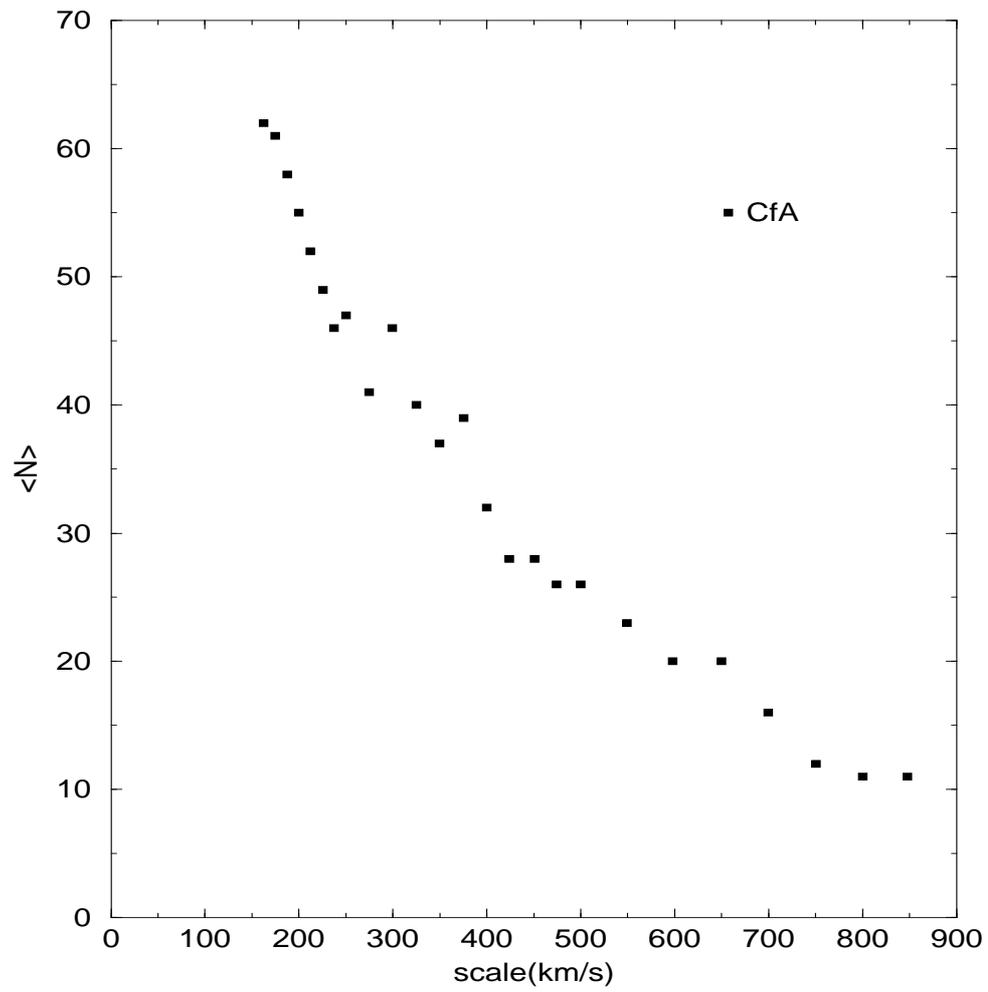

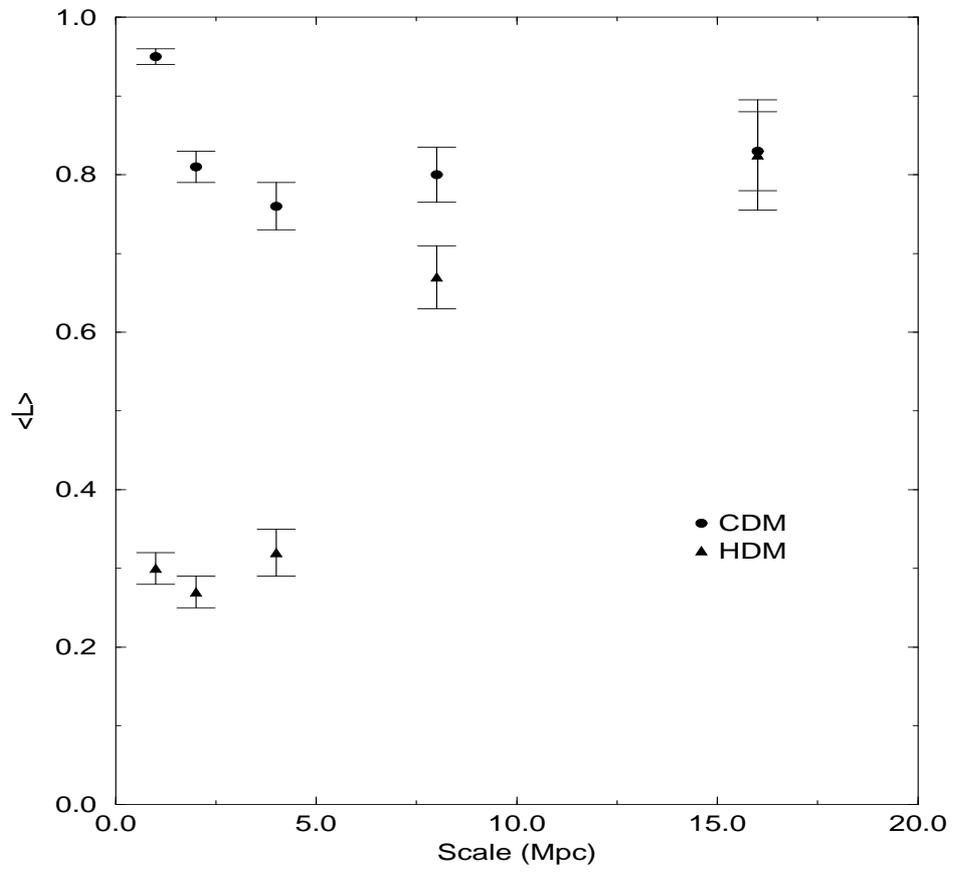

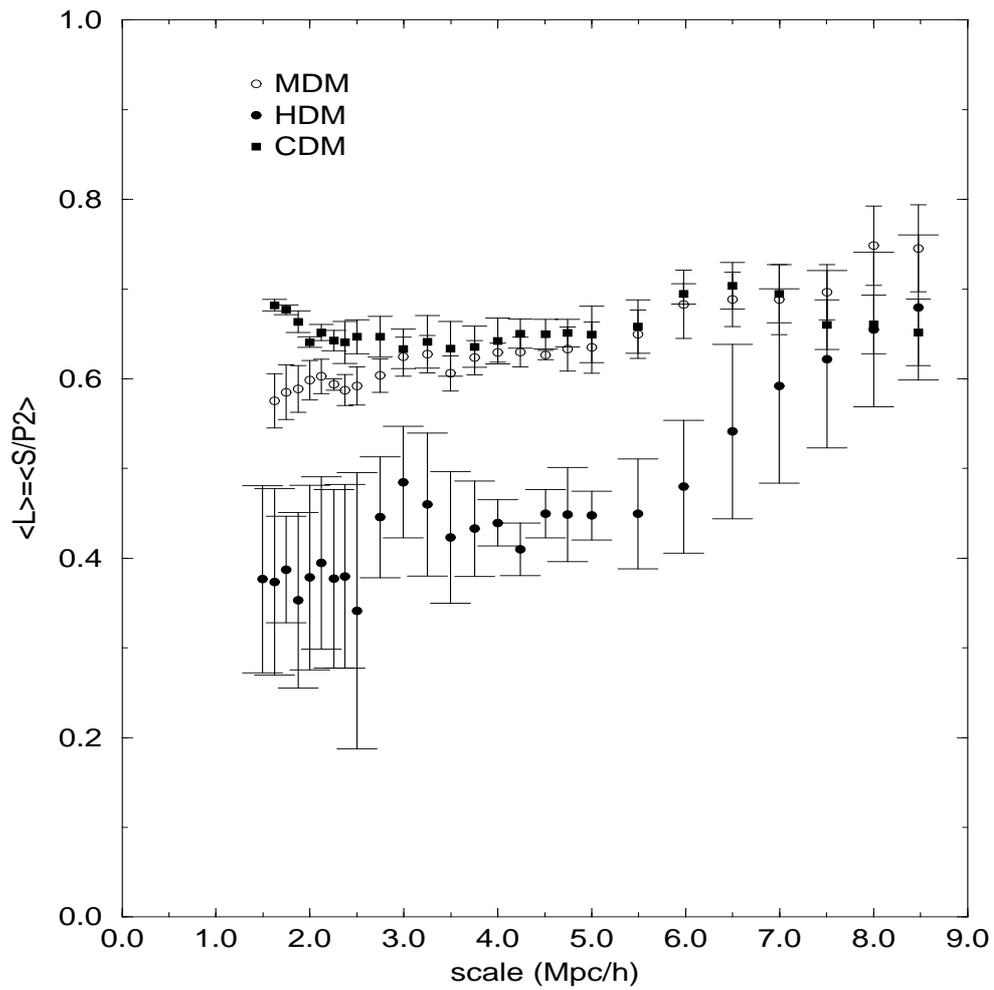

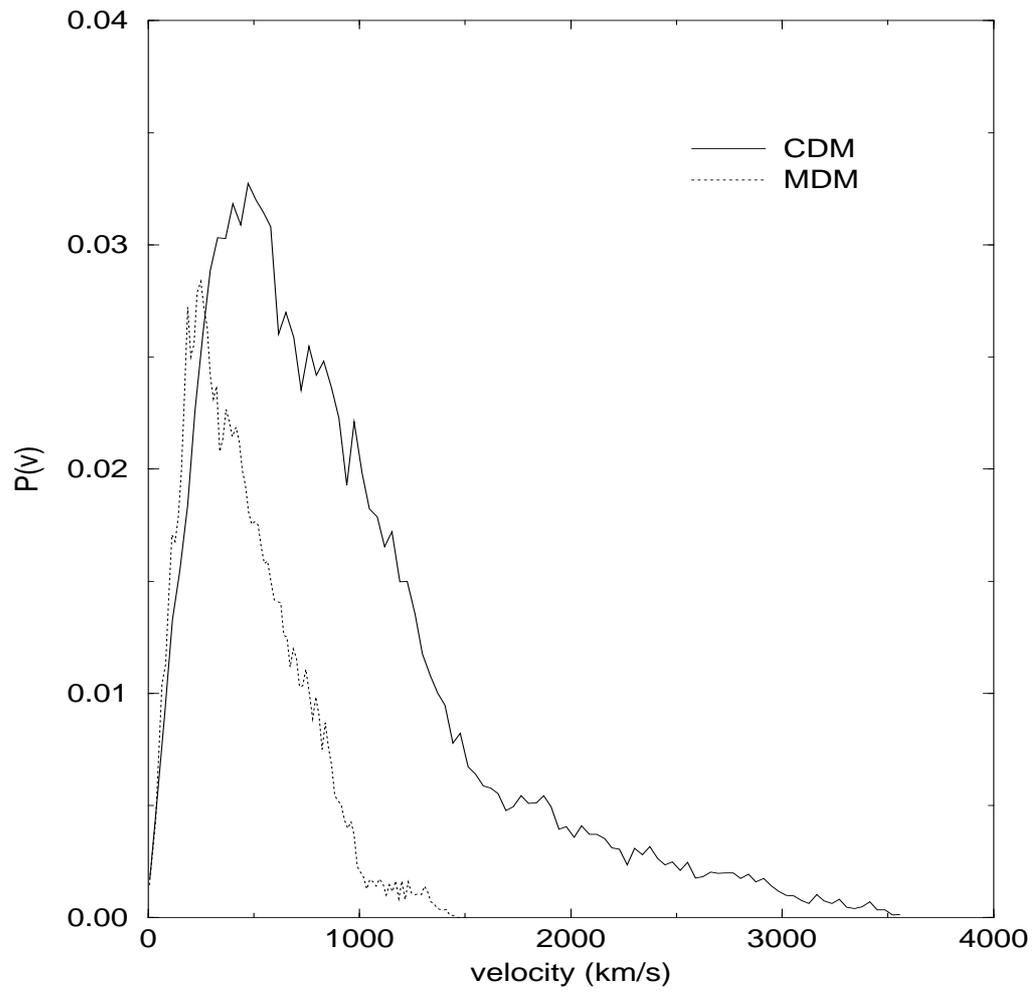

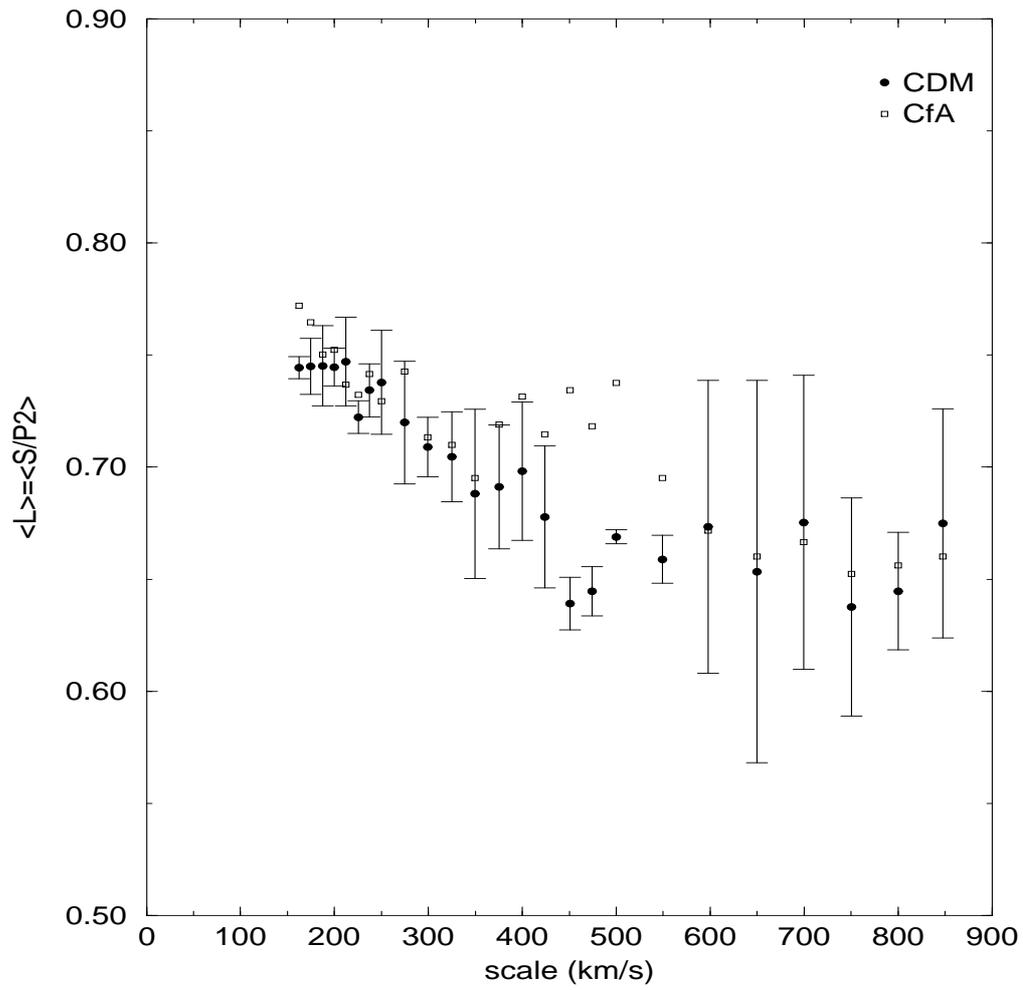

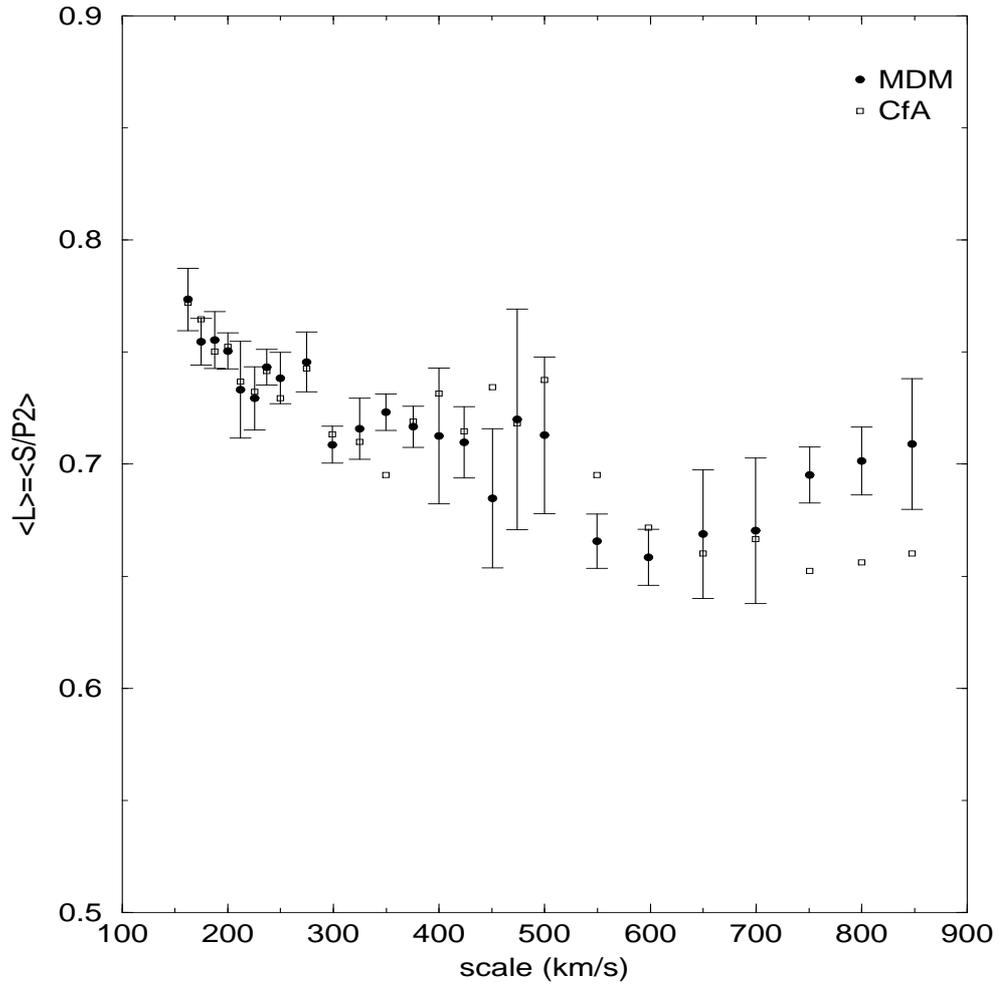